\documentclass[twocolumn,10pt]{asme2e}

\usepackage{multirow}
\usepackage{multicol}
\usepackage{graphicx}
\usepackage{float}
\usepackage{dblfloatfix}
\usepackage{placeins}
\newcommand\Tstrut{\rule{0pt}{2.6ex}}         
\usepackage[justification=centering]{caption}
\usepackage{array,booktabs}

\confshortname{IMECE 2021}
\conffullname{the ASME 2021 \\
            International Mechanical Engineering Congress and Exposition}

\confdate{1-5}
\confmonth{November}
\confyear{2021}
\confcity{Virtual}
\confcountry{Online}

\papernum{IMECE2021-69923}

\title{A COMPARATIVE STUDY OF VARIOUS DEEP LEARNING TECHNIQUES FOR SPATIO-TEMPORAL SUPER-RESOLUTION RECONSTRUCTION OF FORCED ISOTROPIC TURBULENT FLOWS}

\author{T.S.Sachin Venkatesh$^1$, Rajat Srivastava$^2$, Pratyush Bhatt$^2$, Prince Tyagi$^2$, Dr. Raj Kumar Singh$^2$\\
\affiliation{$^1$Department of Applied Physics, Delhi Technological University, New Delhi, India \\
$^2$Department of Mechanical Engineering, Delhi Technological University, New Delhi, India}
}

\begin{document}

\maketitle    

\begin{abstract}
{\it Super-resolution is an innovative technique that upscales the resolution of an image or a video and thus enables us to reconstruct high-fidelity images from low-resolution data. This study performs super-resolution analysis on turbulent flow fields spatially and temporally using various state-of-the-art machine learning techniques like ESPCN, ESRGAN and TecoGAN to reconstruct high-resolution flow fields from low-resolution flow field data, especially keeping in mind the need for low resource consumption and rapid results production/verification. The dataset used for this study is extracted from the ‘isotropic 1024 coarse’ dataset which is a part of Johns Hopkins Turbulence Databases (JHTDB). We have utilized pre-trained models and fine tuned them to our needs, so as to minimize the computational resources and the time required for the implementation of the super-resolution models. The advantages presented by this method far exceed the expectations and the outcomes of regular single structure models. The results obtained through these models are then compared using MSE, PSNR, SAM, VIF and SCC metrics in order to evaluate the upscaled results, find the balance between computational power and output quality, and then identify the most accurate and efficient model for spatial and temporal super-resolution of turbulent flow fields.\\}
\end{abstract}

Keywords: super-resolution, transfer learning, isotropic turbulent flow fields

\section{INTRODUCTION}

In the recent years that have gone by, deep learning models have been intensively used to tackle numerous problems in fluid mechanics. There is an exponential increment in the amount of data available through experiments and simulations due to the rapid improvement in the performance of computational architectures and advances in experimental measurement capabilities. Machine learning provides a wealth of techniques that can learn from this data to solve the prominent challenges and problems in fluid dynamics such as experimental data processing, reduced-order modeling, optimization and turbulence closure modeling. Convolutional Neural Networks serve as a powerful tool for the extraction of fluid dynamic features and thus aid in the prediction of flow fields. Thus, various models have been developed based on CNN to refine the results obtained through CFD analysis quickly with remarkable accuracy.
 
Super-resolution is an important set of image processing techniques that enhances or upscales the resolution of an image or video. These methods thus enable us to reconstruct high-fidelity images from low-resolution ones. In the past few years, with the accelerated growth of techniques of deep learning, various deep learning-based super-resolution models ranging from the classic Convolutional Neural Networks (CNN) to more recent and promising Generative Adversarial Networks (GAN) have been explored and developed which often achieve state-of-the-art performance. It is also to be noted that we undertook this endeavor to not actually solve flow fields or predict them, there are already many out there which are state of the art, but instead we focused on the visualization of the flow field itself which is very often the problem associated with low resource simulations. These techniques when employed on flow fields enabled us in reconstruction of the high-resolution flow (HR) fields from the low- resolution (LR) fluid flow data, thus saving an enormous amount of time and computational power to perform the high-resolution flow field simulations.

Transfer learning is a technique that utilizes pre-trained models and allows us to mix them too, to minimize the computational resources and the time required for the implementation of the super-resolution models. Thus, transfer learning allows models that have been conditioned for one task to be repurposed for another particular task. The Develop model approach and the Pre-trained methodology are the two most widely used methods in transfer learning. In the Develop model approach, a similar predictive modeling problem with plenty of data is solved and any relationship between input and output data is selected, and then a skillful model is developed, trained, and tested for the first problem. This model is then fit on the source task is utilized as the starting point for the model which we need to develop to derive a solution for the problem of interest. This involves either using the entire model or parts of the model, depending on the problem and the modeling technique used. On the basis of the input-output pair data required for the task of interest, the model may also be customised or optimised. In a pre-trained model approach, a pre-trained source model is chosen from available models already trained on large and challenging datasets released by various research institutions. Then, this model acts as an initial point for development of the model specific to the problem statement. Either the entire model or parts of it are utilized and tuned to achieve the desired results.

\section{LITERATURE REVIEW}

 In 2019, a data-driven approach was presented by Sekar, Qinghua, Chang, and Khoo for the prediction of incompressible laminar steady flow fields over airfoils developed by combining CNN and deep Multilayer Perceptron (MLP) \cite{2019PhFl...31e7103S}. Subgrid modeling assisted by machine learning for large eddy simulation (LES) was proposed by Maulik in 2019 \cite{PhysRevE.97.042322}. Combined CNN and Long short-term memory (LSTM) were used by Hasegawa for developing a reduced-order model (ROM) for a two-dimensional unsteady wake behind various bluff bodies in 2020 \cite{2020ThCFD.tmp...26H}. In 2019, Fukami, Fukagata, and Taira \cite{2019JFM...870..106F} performed super-resolution to reconstruct the high-resolution flow field of homogeneous decaying isotropic turbulence on low resolution turbulent flow field data using CNN and hybrid down sampling skip multiscale (DSC/MS) models. The multiscale model showed remarkable accuracy in the generation of HR flow fields from LR snapshots. Then in 2020, they performed super-resolution analysis based on DSC/MS machine learning model using spatial super-resolution and temporal inbetweening techniques which was able to recover high-resolution turbulent flow fields from low resolution or coarse flow data in space and time \cite{2021JFM...909A...9F}. In 2019, Deng, He, Liu, Kim \cite{2019PhFl...31l5111D} utilized advanced neural networks like super-resolution generative adversarial networks (SRGAN) and enhanced-SRGAN (ESRGAN) to enhance the spatial resolution of turbulent flow behind two side by side cylinders. Various other deep learning models such as static convolutional neural network (SCNN) and Multiple temporal path convolutional network (MTPC) have also been utilized to fully capture features in different time ranges  by Liu, Tang, Huanga and Lu \cite{2020PhFl...32b5105L}. Techniques for super-resolution and denoising of fluid flow using a physics informed CNN without using any HR labels for training have also been demonstrated by Gao, Sun, Wang \cite{2020arXiv201102364G}. Several flow SR problems relevant to cardiovascular applications were proposed in the study to evaluate the method's effectiveness and merit. In 2020, Stengel, Glaws, Hetinger, and King developed an adversarial deep learning model to super resolve the atmospheric turbulent winds and solar outputs from global climate models up to fifty times \cite{2020PNAS..11716805S}. Xu, Luo, Wang and You \cite{2020ApOpt..59.5729X} used 3D super-resolution generative adversarial network (3D-SR-GAN) so as to develop a data-driven 3D super-resolution approach which can enhance the spatial resolution of a three dimensional turbulent jet flame structure by two times in all the spatial direction \cite{EasyChair:5975}. Meta-Transfer Learning for Zero-Shot Super-Resolution (MZSR) \cite{2020arXiv200212213S}, is a technique that finds a generic initial parameter which is suitable for internal learning has been proposed by Sho in 2020. This method thus allows the network to quickly adapt to a given image condition. In 2019, Zhang, Wang, Lin and Qi designed an end-to-end deep learning model \cite{2019arXiv190300834Z} which is able to enrich the high-resolution details of the image by adaptive transfer of the the texture from the reference images according to their textural similarity by conducting a multilevel matching in the neural space.

\section{MATERIALS AND METHODS}

\subsection{Super resolution}

Super-resolution is an approach to improve the resolution/quality of both spatial and temporal data. An image or video might be in Low-Resolution form either due to downsampling or degradation. The mathematical function $LR=(Degrade)HR$ states the relation between the two aspects of the data. We here observe that the Low Resolution data can be obtained by down-sampling the High Resolution data. However here the objective is on the contrary. We have to obtain the HR from the LR and that too in the absence of the downsampling/degrading function. In recent years deep learning methodologies have been proven effective in obtaining the HR i.e., High-Resolution image from the LR i.e., Low Resolution or degraded image. Deep learning models and techniques for super-resolution utilized in our study are explained in the subsequent sections.   

\subsubsection{ESPCN}\hfill

ESPCN has an Efficient Sub Pixel convolutional layer added to the CNN which enhances the resolution towards the end of the network. The network of ESPCN can be represented as an efficient sub-pixel convolutional layer \cite{shi2016realtime} that includes a convolution having a fractional stride of $1/r$ in LR space with $W_s$ as a filter having size $K_s$. Weights that are not in alignment with the pixels are to be omitted for calculations. This results in the number of activation patterns to be $r^2$. These activated patterns have the greatest number of activated weights in accordance with the location. 

\begin{figure}[htp!]
    \centering
        \includegraphics[width=1\linewidth]{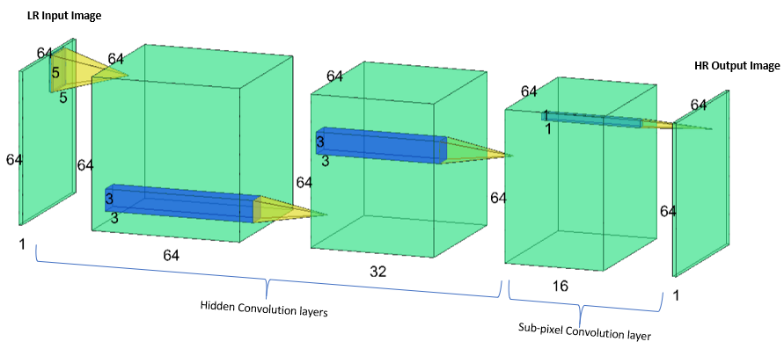}
    \caption{ESPCN network architecture}
\end{figure}
\vspace{-25pt}
\begin{equation}
I^{S R}=f^{L}\left(I^{L R}\right)=P S\left(\mathrm{~W}_{L} * f^{L-1}\left(I^{L R}\right)\right)+\mathrm{b}_{L}
\end{equation}
where PS is the periodic shuffling operator and $W_L$, $b_L$ are the weights and biases respectively. The PS or periodic shuffling operator shuffles the elements of the tensor. In certain deconvolution procedures, we pad the images with zeros until convolution, which may have a negative impact on the outcome. While performing pixel shuffle at the network's final layer to recover the LR image, no padding operation is needed. The above equation can be mathematically represented as:
\begin{equation}
{PS}({T})_{x, y, c}={T}_{\left[\frac{x}{r}\right] \cdot\left[\frac{y}{r}\right], {c.r.mod}({y}, {r})+{c.mod}({x}, {r})+{c}}
\end{equation}

Thus, $W_L$ having a shape of $n_{L-1}*r^2*C*k_L*k_L$, this new layer together with CNN layers is termed as Efficient Sub Pixel Convolutional Network or ESPCN. In this paper, the Low-Resolution videos are generated by downsampling the High-Resolution Data so the actual High-Resolution data will be useful in generating the MSE or Mean Squared Error. This MSE is used to calculate the difference between the Super-Resolution Videos generated through the network and the actual High-Resolution Videos. The loss function in ESPCN is given by
\begin{equation}
\mathrm{l}\left(\mathrm{W}_{1: L}, b_{1: L}\right)=\frac{1}{r^{2} H W} \sum_{x=0}^{r H} \sum_{y=0}^{r H}\left(I_{x, y}^{H R}-f_{x, y}^{L}\left(I^{L R}\right)\right)^{2}
\end{equation}

\subsubsection{ESRGAN} \hfill

Enhanced Super-Resolution Generative Adversarial Networks better known as ESRGAN,  as the name implies is an upgraded SRGAN \cite{DBLP:journals/corr/LedigTHCATTWS16} which is a super-resolution technique for single images based on generative adversarial network . It employs a perceptual loss mechanism that includes an adversarial loss and a material loss. In ESRGAN the network's overall high-level architecture nature is preserved, but a few new ideas are introduced and updated, resulting in a more effective network. The Residual-in-Residual Dense Block (RRDB) is introduced into the Generator network configuration, which increases the network's power and makes training simpler. 

\begin{figure}[htp!]
    \centering
        \includegraphics[width=1\linewidth]{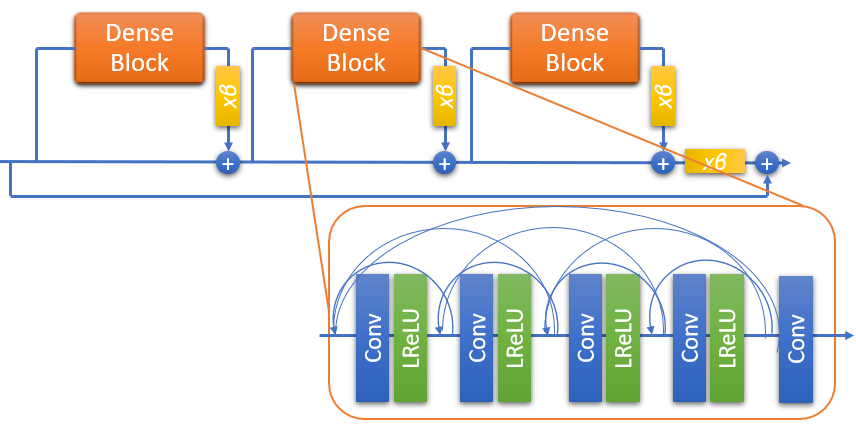}
    \caption{Residual in Residual Block (RRDB) of ESRGAN}
\end{figure}

The ESRGAN model retains the SRGAN high-level architecture framework and uses the new basic component called RRDB \cite{DBLP:journals/corr/abs-1809-00219}. The proposed RRDB structure is more complex and diverse as compared to the initial residual structure in SRGAN. In addition, to the improved design, it employs a number of techniques to facilitate the preparation of a very profound network: 1) Residual scaling, which is the method of decreasing residuals by proliferating them by a constant ranging from 0 to 1 prior addition to the main path to avert the erratic behavior;  2) Decreased initialization, since it was found factually that remnant structure is simpler to train as the initial parameter variance decreases.

In addition to the traditional discriminator, ESRGAN employs the relativistic GAN, which attempts to estimate the likelihood that the actual picture is more realistic than a false image.
\begin{equation}
\begin{array}{l}
{D}_{R a}\left({x}_{r}, {x}_{f}\right)=\sigma\left(C\left({I}_{r}\right)-{E}\left[C\left({I}_{f}\right)\right] \rightarrow 1\right.\\
\\
{D}_{R a}\left({x}_{f}, {x}_{r}\right)=\sigma\left(C\left({I}_{f}\right)-{E}\left[C\left({I}_{r}\right)\right] \rightarrow 0\right.
\end{array}
\end{equation}

where $D_{Ra}$ is Relativistic Average Discriminator, $I_r$  and $I_f$ denote Real image and Fake image respectively and E denotes the operation of taking an average of all data in mini-batch. We build a more effective perceptual loss $L_{percep}$ by controlling features before activation rather than after, as SRGAN does. Using features after they have been enabled causes differences in recreated brightness as compared to the ground-truth picture.
\vspace{-3pt}
\begin{equation}
{L}_{G}=={L}_{percep}+\lambda{L}_{G}^{Ra}+\eta{L}_{1}
\end{equation}
Where $L_1$ represents the loss that computes the 1-norm distance from reconstructed image to ground truth, $\lambda$, $\eta$ are the loss term balancing coefficients.

\subsubsection{TecoGAN}\hfill

GANs (generative adversarial networks) have shown to be remarkably efficient at solving complicated distributions. Explicitly applying GANs to sequence generation without carefully designed restrictions, on the other hand, usually results in strong artifacts over time. Conditional video generation functions, in particular, have extreme learning difficulties. It presents a recurrent adversarial generative model for VSR that consists of three structures: a recurrent generator, a flow estimation network, and a spatial-temporal discriminator. The generator $G$, is used to generate significant video frames from low-resolution inputs frequently \cite{Chu_2020}. The flow estimation network $F$, learns flow adjustment between frames to support both the generator and the spatial-temporal model.
\begin{equation}
\mathrm{v}_{t}=\mathrm{F}\left(\mathrm{i}_{t-1}, \mathrm{i}_{t}\right), \mathrm{g}_{t}=G\left(\mathrm{a}_{t}, W\left(\mathrm{~g}_{t-1}, \mathrm{v}_{t}\right)\right)
\end{equation}

The $g_T$, high-resolution Output, is generated by the generator from the LR data input $x_T$ and then sequentially uses the previously produced HR output $g_{T-1}$. The architecture of the generator can be represented as:

\begin{figure}[htp!]
    \centering
        \includegraphics[width=1\linewidth]{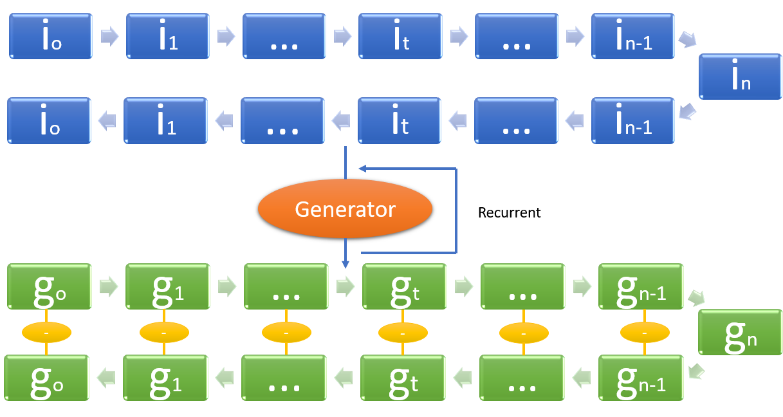}
    \caption{Frame Recurrent Generator for VSR }
\end{figure}

In this case, the network F has been trained to evaluate the motion $v_t$ from frame $i_{t-1}$ to $i_t$ and W stands for warping.

\begin{figure}[htp!]
    \centering
        \includegraphics[width=1\linewidth]{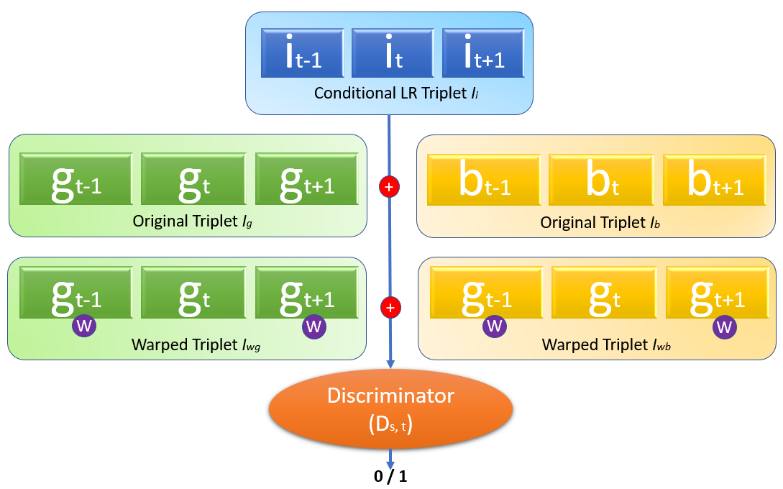}
    \caption{Conditional Discriminator for VSR }
\end{figure}

The architecture of the discriminator can be seen in the Fig 4 \cite{DBLP:journals/corr/abs-1811-09393}. This sequence of the model is passed with two input sets: i) the actual structure, ii) the generated structure. The shape of both sets is the same, having equal numbers of High-resolution data and corresponding Low-resolution frames. The actual structure has the high resolution frames with natural data and on the other hand, the generated Structure is made up of the data generated by the $G$. The objective of having two input sets is that the model can be penalized by the Discriminator for the structures generated having less spatial-temporal details or vice versa. 

\subsection{Transfer Learning}
One of the major challenges faced while training a neural network is to figure out the exact amount of the dataset and also the quality of the dataset. Training large models can be very difficult as it not only requires a vast amount of data set but also enormous computation power. Transfer learning provides a good advantage in overcoming these challenges. It is the repeated usage of the pre-trained model for a new problem statement similar to the previous one \cite{10.5555/1803899}. The intention in transfer learning is to save the information that a model has acquired while solving a problem with enormous amounts of labeled data and use it to resolve another similar problem with very little data available to train. Rather than starting from scratch, it starts from a pattern having similarities with the sample solved beforehand. In this, we prevent the first few layers from training. The first few layers are common in general. The number of layers to be frozen depends upon the similarity of the problem. We only need to put in new layers for the unrecognized features. In case of large inaccuracy, the layers can be retrained from start. The application of transfer learning can be easily seen in computer vision, NLP as they require huge amounts of data to train on. Transfer Learning not only saves time as well as saves us from the complicated computations which require a heavy usage setup.

\subsection{Metrics}
\subsubsection{Mean Squared Error (MSE)} \hfill

Mean Squared Error commonly called MSE is one of the widely used loss functions used for calculating the mean squared error of each pixel between two comparative images. The MSE \cite{995823} between two images I (x, y) and I' (x, y) can be depicted as:
\begin{equation}
M S E=\frac{1}{M N}+\sum_{x=1}^{M-1} \sum_{y=1}^{N-1}\left[I^{\prime}(x, y)-I(x, y)\right]^{2}
\end{equation}
A lower value of MSE means both images are quite identical and vice versa. For example, an MSE of 0 represents both images as identical. The only concern with MSE is that it is extremely dependent on image intensity scaling.

\subsubsection{Peak Signal to Noise Ratio (PSNR)} \hfill

Peak Signal-to-Noise Ratio is a comparison and benchmarking metric, calculated as the ratio of the highest potential power of a signal to the power of unwanted noise that reduces the accuracy of its representation. It is most widely used to assess the accuracy of lossy compression codec restoration. PSNR is measured in logarithmic values with decibel(dB) as a reference scale\cite{HuynhThu2008ScopeOV} and mostly uses MSE for its computation:
\begin{equation}
\begin{array}{c}
P S N R=10 \log _{10} \frac{\mathrm{MAX}_{I}^{2}}{M S E} = 20 \log 10\left(\mathrm{MAX}_{I}\right)-10 \log 10(\mathrm{MSE})
\end{array}
\end{equation}
Here, the maximum possible pixel value of the image is given by $MAX_I$. Conventional PSNR values in highly compressed images and video compressions are from 30 and go upto 50 dB, with pixel value of 8 bits, with higher being stronger. Thus, a higher value signifies that the images are more detailed.

\subsubsection{Spectral Angle Mapper (SAM)} \hfill

A pixel vector in N-dimensional space of spectra has both magnitude and an angle which is measured from the respected coordinate axis. Spectral Angle Mapper (SAM) is a spectral classification that matches pixels to reference spectra using only the n-D angular component of the pixel vector so when comparing the two spectra the multidimensional vector is defined for both and the angle formed between them is calculated. Smaller angles indicate further similarity to the reference continuum. Both spectra are treated as matched if the angle is below a threshold value and unclassified if the angle is further away from maximum threshold. 

\subsubsection{Visual Information Fidelity (VIF)} \hfill

The Visual Information Fidelity (VIF) index \cite{1532311} is a comprehensive image caliber evaluation index that has grounds on  natural scene statistics and the definition of image data derived from the human visual system. To measure the data exchanged among the trial and reference images, the VIF indicator engages natural scene statistical (NSS) models in combination with a distortion (channel) model. Furthermore, the VIF index is founded on the conjecture that this mutual knowledge is an allegiant feature that is closely in relation with the visual consistency. A higher VIF index is obtained when the images are almost similar. In comparison to antecedent approaches focusing on human visual system (HVS) responsiveness to error and composition analysis, this computational perspective employed in an informative theory context produces a full reference (FR) quality evaluation (QA) technique that is independent of any HVS or observing geometrical variable, nor any constants that need to be optimised, thus remaining competitive with cutting-edge QA methods.

\begin{figure*}[t]
  \includegraphics[width=1\textwidth]{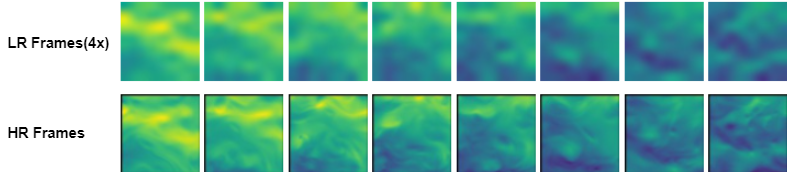}
  \caption{4x degraded LR video frames (top) of forced isotropic turbulence and their corresponding HR video frames (bottom)}
\end{figure*}

\subsubsection{Spatial Correlation Coefficient (SCC)} \hfill

Spatial correlation is a statistical dependency of spatial direction between two signals. It means there is a relationship between the average signal gain obtained and the angle of arrival of a signal. Natural images have a high degree of spatial similarity but real-world situations, the correlation will be lower, but high odds will also be exceeded.  This correlation is directly dependent on the eigenvalues of matrices which are in correlation. Higher the correlation coefficient higher the similarity in images. The eigenvectors demonstrate the spatial direction of the matrix.  High spatial correlation represents that the eigenvalue spread is sizeable and vice versa. 
    
\begin{figure*}[b]
\centering
  \includegraphics[width=0.8\textwidth]{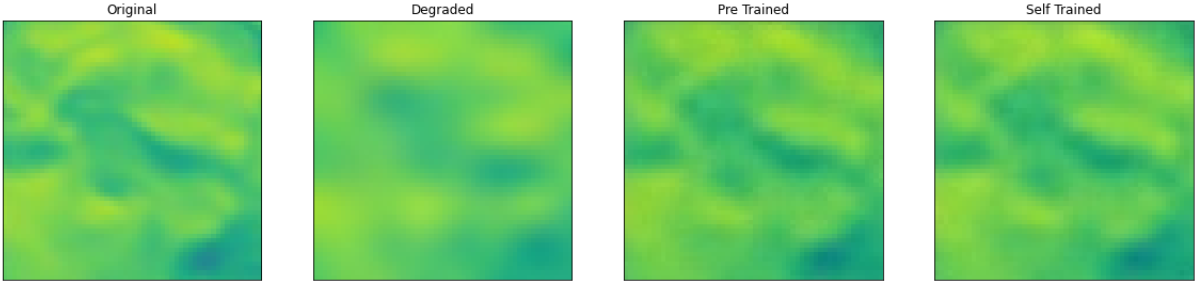}
  \caption{ESPCN results}
\end{figure*}

\section{DATASET}
The dataset contains 100 high resolution (HR) videos and corresponding 100 4x degraded low resolution (LR) videos of forced isotropic turbulent flow fields. Each video has a duration of 8s and dimensions 64 x 64 pixels. Each high-resolution video was obtained by stacking 200 timestamps of forced isotropic flow fields obtained by plotting pressure for 64x64 points. Consecutive timestamps are separated by a time step of 0.02s. The HR videos are converted into LR by first downscaling and then upscaling the HR snapshots by 4 times using bicubic interpolation and then stacking these 200 snapshots to obtain the LR video.  The HR 64x64 snapshots or timestamps are obtained from the ‘isotropic 1024 coarse’ dataset which is a part of Johns Hopkins Turbulence Databases (JHTDB). As stated in the description of the database, the data was obtained by direct numerical simulation (DNS) of forced isotropic turbulence on a 1024$^3$ periodic grid, with viscosity($\nu$) equal to 0.000185 and Taylor Reynolds number ($R_\lambda$) fluctuating around 433, by utilizing a pseudo-spectral parallel code.

\begin{figure*}[t]
\centering
  \includegraphics[width=0.8\textwidth]{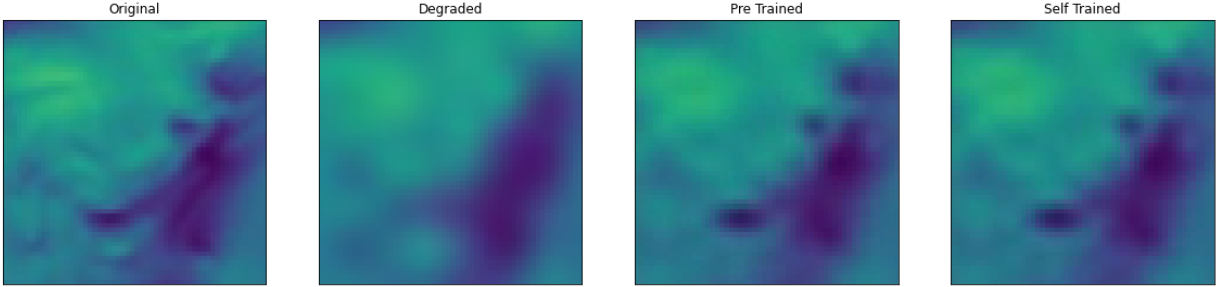}
  \caption{ESRGAN results}
\end{figure*}

\begin{figure*}[b]
\centering
  \includegraphics[width=0.8\textwidth]{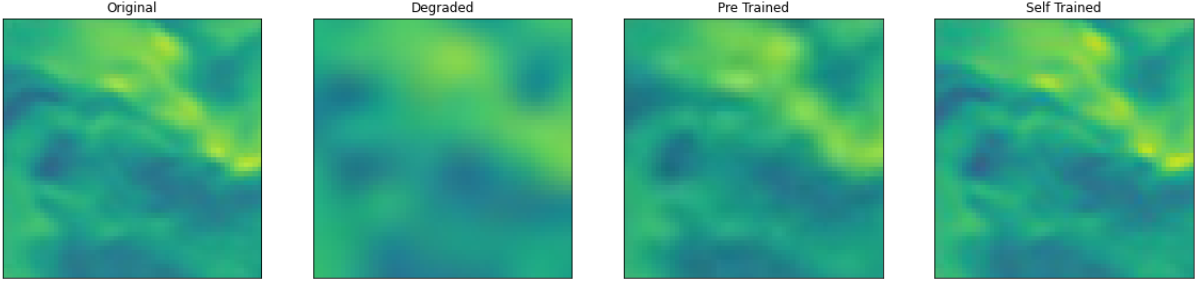}
  \caption{TecoGAN results}
\end{figure*}

Analyzing the time convergence of the viscous term was accomplished using an integrating factor. The nonlinear term is written in vorticity form, and the remaining terms are combined using a second-order Adams-Bashforth scheme. After the simulation reaches a statistically stationary state, 5028 frames of data, which consists of pressure and the 3 components of the velocity vector, are generated and inserted inside the database. The entire duration of the stored data is about five large-eddy turnover times. The isotropic coarse dataset thus possesses 5028 timesteps, for time t between 0 and 10.056 s. The dataset of 100 HR and 100 LR videos is segmented in the ratio of 80:10:10 for training, validation, and testing respectively.

\section{RESULTS AND DISCUSSION}
We performed exhaustive evaluations of different models on our dataset with different techniques of architecture handling and model training. As a preliminary test of the model’s capabilities, we  use the pre-trained weights to obtain results and use them as a benchmark. We then use transfer learning on these models to partially train and tweak them. Results obtained through the use of both these methods are discussed for the various models used in our study in the following section. To ensure uniformity in results and performance, we employ an adaptive data rate with an initial value of 0.01 which scales by 10\%. We also implement the Adam optimizer in all models as it synergizes well with the adaptive learning rate and also works best in low memory/computing power scenarios.

\subsection{ESPCN}
We chose this model because of its ease of use and a comparatively simple architecture. Since it is based on Convolutional Neural Networks, albeit an advanced one, its architecture is straightforward and relies mostly on its Sub Pixel Convolution layer to do most of the work. Therefore the pre trained model which was trained on the CDVL database consisting of videos set in everyday scenarios performs a poor job at upscaling our turbulence images. The results generated although bring back the pressure differences, aren’t very clear in defining their boundaries and the depth. 

We then froze the first two layers of the model and trained the final Sub Pixel layer on our dataset for 10 epochs, each with a batch size of 80. This rapid training produced highly detailed images as compared to the ones produced by the pre trained model. After some fine tuning, the produced images could bring back a some lost details like the enhanced colour variance, but it still couldn’t produce a concrete result as the output images lacked detail in areas with rapid change in pressure. We also encountered random artifacts as the model hit overfitting in less than 15 epochs which inversely affected the image quality and the model statistics. 

\subsection{ESRGAN}
ESRGAN produces commendable results particularly with respect to our dataset. It utilizes its Generative Adversarial Network to upscale images to satisfaction, but shows minimal improvement when it is partially trained. It is able to differentiate between high and low pressure regions and upscales them as whole blocks but the model compensates for it by averaging the neighboring pixels in areas of minimal change defeats the purpose of the model.

To fully harness the capability of this model, we trained different layers partially and found that freezing all the layers except for the last layer of the generator and the whole discriminator produces the best results in an efficient manner. Since the generator is constantly being updated with counter results from the discriminator, we train the discriminator to classify our dataset. But this resulted in minimal to average improvement depending on different images.

This is visible in the results presented in Fig 7, where both the pre trained model and the self trained model are able to identify areas of differing pressure, highlight it and upscale it but the pre trained model leaves out the minute intricacies such as the waves and curves. In comparison, the self trained model does all that the pre trained model does but also account for the small scale changes like the curves and waves that the pre trained model left out.

The other caveat of this model is that it is trained to super resolve images and not videos. Therefore, it does not account for temporal coherence and instead trades it for enhancing the quality of a single image, one at a time. Thus, even though ESRGAN scores well in the single image criteria, it loses out in preserving the flow.

\begin{figure*}[t]
    \centering
    \begin{minipage}{0.5\textwidth}
        \centering
        \includegraphics[width=1.07\linewidth]{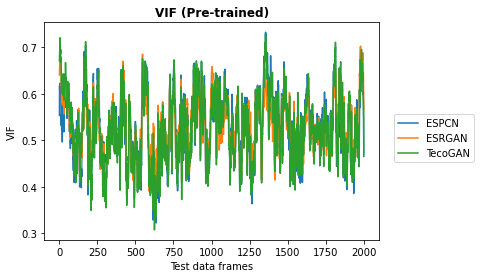}
    \end{minipage}%
    \begin{minipage}{0.5\textwidth}
        \centering
        \includegraphics[width=0.87\linewidth]{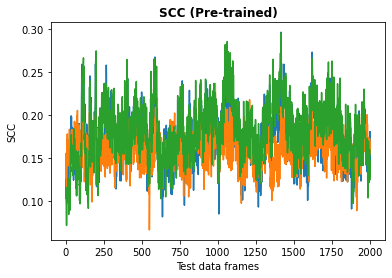}
    \end{minipage}
    \caption{Frame to frame statistics of VIF and SCC metrics of results generated by pre trained models}
    \begin{minipage}{0.5\textwidth}
        \centering
        \includegraphics[width=1.07\linewidth]{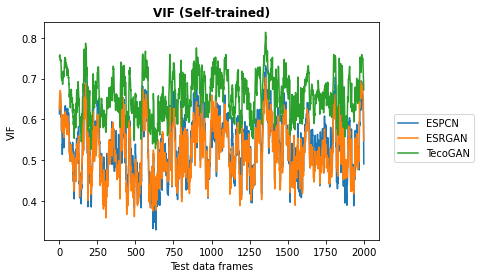}
    \end{minipage}%
    \begin{minipage}{0.5\textwidth}
        \centering
        \includegraphics[width=0.87\linewidth]{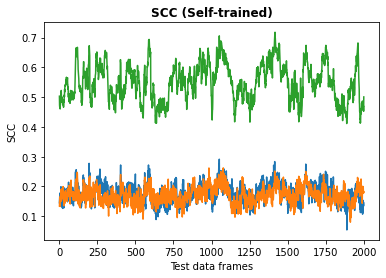}
    \end{minipage}
    \caption{Frame to frame statistics of VIF and SCC metrics of results generated by self trained models}
\end{figure*}

\subsection{TecoGAN}
Of all the different models we examined and shortlisted, TecoGan performs extraordinarily well even with limited amount of resources and computational power. Since a heavy emphasis is laid on temporal coherency, TecoGAN learns from the accompanying frames of the dataset to learn the changes in pressure with time, therefore it not only enhances the low resolution data but also preserves the ‘flow’ of the video which other models don’t. 

It compensates for the data loss in low pressure regions or the ‘dips’ by examining the near future frames to  produce those details while upscaling LR images. The pre trained model of TecoGAN was really well trained as it could produce above average results out of the box as compared to the results from other pre trained models. 
Since the pre trained model was already capable enough, we only had to fine tune the generator, thus we focused on training the discriminator similar to ESRGAN.

The self trained model produced high resolution images which were almost as good as the ground truth and were very similar to their corresponding original images. This is evident by the value of MSE for TecoGAN's self trained version. The pre trained model was able to reconstruct the features but they were still not properly defined but the self trained model was able to exactly replicate the features right to the minute changes while preserving the boundary. The boundary problem is of prime importance as it may highly distort/skew the pressure field. This added advantage of being able to exactly replicate the image and its features complements TecoGAN's flow preservation very well and produces a highly synergic model.

\begin{table*}[t]
\begin{center}
\begin{tabular}{||c||c||c||c||c||c||c||c||}
\cline{1-8}
\multirow{2}{*}{\textbf{Metrics}} & \multirow{2}{*}{\textbf{Degraded}} & \multicolumn{2}{c||}{\textbf{ESPCN}}         & \multicolumn{2}{c||}{\textbf{ESRGAN}}          & \multicolumn{2}{c||}{\textbf{TecoGAN}}       \\ 
\cline{3-8}
                                  &                                    & \textbf{Pre-Trained} & \textbf{Self-Trained} & \textbf{Pre-Trained} & \textbf{Self-Trained} & \textbf{Pre-Trained} & \textbf{Self-Trained}  \\ 
\hline
\textbf{MSE}                      & 30.9911                            & 26.8544              & 21.3932               & 16.1058               & 14.0273               & 21.8246              & 5.5169                  \\
\textbf{PSNR}                     & 34.3949                            & 34.5287              & 34.8281               & 36.4730              & 37.0000               & 35.0206              & 41.2876                \\
\textbf{SAM}                      & 0.0598                             & 0.0443               & 0.0424                & 0.0380               & 0.0365                & 0.0387               & 0.0277                 \\
\textbf{VIF}                     & 0.3593                             & 0.5272               & 0.5321                & 0.5275               & 0.5283                & 0.5289               & 0.6610                 \\
\textbf{SCC}                      & 0.0761                             & 0.1745               & 0.1790                 & 0.1619               & 0.1714                & 0.1833                & 0.5506               \\
\hline
\end{tabular}
\caption{Median metrics of 10 videos (test set) of the dataset obtained from the pre and self trained models of ESPCN, ESRGAN and TecoGAN}
\end{center}
\end{table*}
\FloatBarrier

\begin{table*}
    \centering
    \begin{tabular}{>{\centering}m{0.2\linewidth} >{\centering}m{0.15\linewidth} >{\centering}m{0.15\linewidth} >{\centering}m{0.15\linewidth} >{\centering\arraybackslash}m{0.2\linewidth}}
        \toprule
        \textbf{Bicubic 4x} & \textbf{ESPCN} & \textbf{ESRGAN} & \textbf{TecoGAN} & \textbf{Ground Truth} \\
        \midrule
        \\[-1em]
        \multirow{2}{*}{\includegraphics[width=0.75\linewidth]{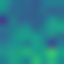}} & \includegraphics[width=1\linewidth]{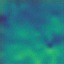} & \includegraphics[width=1\linewidth]{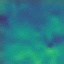} & \includegraphics[width=1\linewidth]{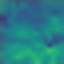} & \multirow{2}{*}{\includegraphics[width=0.75\linewidth]{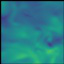}} \\
         & \includegraphics[width=1\linewidth]{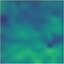} & \includegraphics[width=1\linewidth]{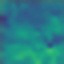} & \includegraphics[width=1\linewidth]{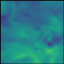} & \Tstrut\\ 
        \midrule
        \\[-1em]
        \multirow{2}{*}{\includegraphics[width=0.75\linewidth]{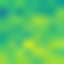}} & \includegraphics[width=1\linewidth]{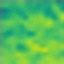} & \includegraphics[width=1\linewidth]{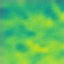} & \includegraphics[width=1\linewidth]{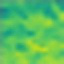} & \multirow{2}{*}{\includegraphics[width=0.75\linewidth]{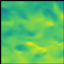}} \\
         & \includegraphics[width=1\linewidth]{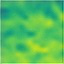} & \includegraphics[width=1\linewidth]{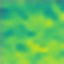} & \includegraphics[width=1\linewidth]{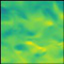} & \Tstrut\\ 
         \midrule
         \\[-1em]
        \multirow{2}{*}{\includegraphics[width=0.75\linewidth]{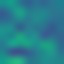}} & \includegraphics[width=1\linewidth]{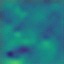} & \includegraphics[width=1\linewidth]{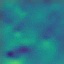} & \includegraphics[width=1\linewidth]{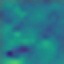} & \multirow{2}{*}{\includegraphics[width=0.75\linewidth]{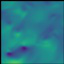}} \\
         & \includegraphics[width=1\linewidth]{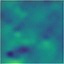} & \includegraphics[width=1\linewidth]{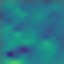} & \includegraphics[width=1\linewidth]{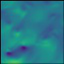} & \\ 
    \bottomrule
    \end{tabular}
    \caption{Visual comparison of some Super-Resolution images generated from different models. The upper sub-row of images are generated from pre trained models while the lower sub-row of images are generated from self trained models}
\end{table*}
\FloatBarrier
\section{CONCLUSION}
Our study was able to conclude that in the low resource settings, transfer learning can be used for super-resolution which is a computationally expensive task and it can be used to recover high-res turbulent flow fields from low-res data, by training only a select number of layers and fine tuning models according to our datatset. TecoGAN produced the best results in terms of VIF for pretrained weights as well as when the model was trained specifically for forced isotropic turbulent flow fields. Thus transfer learning incorporated super resolution can enhance the quality of low-res simulations while making a slight trade-off in the simulation quality for a great reduction in the required computational time and resources for processing. All layers of the model can be further trained to achieve better results in high resource settings. 

\begin{acknowledgment}
The authors would like to thank the Fluid Systems Laboratory of Delhi Technological University for providing computing facilities. We are thankful to our professor Dr. Raj Kumar Singh who provided expertise which greatly assisted us in the research and  improved the paper significantly. We would also like to thank other members of the Fluid Mechanics Group for directly or indirectly helping during the research work.
\end{acknowledgment}

\bibliographystyle{asmems4}
\bibliography{asme2e}

\end{document}